\documentclass[twocolumn,showpacs,preprintnumbers,amsmath,amssymb]{revtex4}

\newcounter{ctr}

\usepackage{graphicx}
\usepackage{epsfig}
\usepackage{dcolumn}
\usepackage{bm}

\begin{document}

\title{On a Condition for Intracellular Adaptive Dynamics
for Chemotaxis}

\author{Masayo Inoue$^1$ and Kunihiko Kaneko$^{1,2}$}
\affiliation{
  $^1$Department of Pure and Applied Sciences, University of Tokyo, \\
  \normalsize 
  3-8-1 Komaba, Meguro-ku, Tokyo 153-8902, Japan \\
  $^2$ ERATO Complex Systems Biology Project, JST, 3-8-1 Komaba,Meguro-ku, Tokyo 153-8902, Japan \\                                }
\date{\today}

\begin{abstract}
Microorganisms often perform chemotaxis,
(i.e., sensing and moving toward a
region with
a higher concentration of an attractive chemical) by changing the rate of tumbling for random walk.
We studied several models
with internal adaptive dynamics numerically to examine the validity
of the condition for chemotaxis proposed by Oosawa and
Nakaoka~\cite{Oosawa}, which states that 
the time scale of tumbling frequency is smaller than that of adaptation and greater than
that of sensing. Suitably renormalizing the timescales showed that the
condition holds for a variety of environments and for both short- and long-term
behavior.
\end{abstract}


\maketitle

\section{Introduction}

Chemotaxis is frequently observed in microorganisms.
Bacteria move in the direction of a higher concentration of an attractive chemical.
Although chemotactic behavior is conserved throughout evolution,
bacteria never move 
toward a goal directly, but instead often tumble by changing direction
randomly. By changing the tumbling frequency, bacteria assemble around
their goal, the region rich
in attractants. Experimental bacterial systems show that
bacteria regulate the rate of tumbling and
their speed, but change direction randomly.
Bacteria move faster but rarely tumble when located in a region with a high concentration
of attractant, whereas they
tumble frequently at low speed in a region with a lower
concentration. However, this experimental observation seems peculiar 
if one considers only normal Brownian movement, because a high tumbling frequency 
should yield a small bacterial diffusion constant 
in the continuum limit, making the attraction to a
high-concentration region impossible.

By assuming an internal state with adaptation dynamics of a certain timescale 
into the element exhibiting Brownian motion,
Oosawa and Nakaoka showed more than 25 years ago that chemotactic behavior is possible~\cite{Oosawa}.
In other words, the existence of memory in the internal state
makes chemotaxis possible.
In contrast, de Gennes~\cite{Gennes} noted recently that the
elements can move in a
favorable direction despite not having an internal state to
retain some memory. However, this is true only for short-term
behavior of chemotaxis. As Clark and
Grant reported~\cite{Clark}, the steady
distribution of long-term behavior is not biased toward a
favorable region.
A temporal change of tumbling frequency with some memory is required to
have a biased steady distribution.
By assuming a class of autocorrelation function for the change of
tumbling frequency, Clark and Grant showed that the steady distribution of
bacteria
can be biased toward the favorable region. In addition, by introducing some
constraints
and optimizing both the
short-term and long-term chemotactic behavior, they obtained a beautiful
solution to
the optimal correlation function.

From a biological viewpoint, the temporal change 
of the tumbling frequency is a result of intracellular
dynamics. For any given intracellular dynamics, arbitrary autocorrelation
function
is not possible, and we were interested in studying the way that specific intracellular
dynamics allow for chemotactic behavior.
Such a condition was proposed previously
by Oosawa and Nakaoka~\cite{Oosawa}, who studied
the steady-state distribution of cells with internal adaptive dynamics 
and the conditions for chemotactic behavior. 
When the environment changes,
bacteria must immediately sense the change and gradually adapt to the new environment
by exploiting the dynamics of the internal state.
Oosawa and Nakaoka showed that bacteria cannot move toward a region with a
higher chemical
concentration if the tumbling timescale $\tau$ is faster than the
sensing timescale $\tau_s$ to
detect the environmental change or slower than the timescale $\tau_a$ for
the adaptation.
When tumbling faster, bacteria tumble randomly without any directional motion,
whereas slower tumbling causes the information about direction to disappear
before the tumbling rate changes.
The proposed condition, which we call the Oosawa condition, states that 
the tumbling timescale $\tau$ is greater
than the sensing timescale ($ \tau_s $) and smaller than the adaptation timescale ($ \tau_a $).
The optimal solution by Clark and Grant~\cite{Clark} satisfies
this Oosawa condition.

In this paper, we introduce
a model of an explicit internal dynamic system with two degrees of
freedom
that responds and adapts to the external environment. These internal dynamics
correspond to some intracellular reaction process, which
exhibits both quick-sensing and slow-adaptive processes where
the parameters $ \tau_s $ and $ \tau_a $ are derived from the
parameters characterizing the intracellular reaction dynamics.
Using this model, we examined the validity of the
Oosawa condition for chemotactic behavior.
We first demonstrate this condition for a chemical concentration field
with a step function
in space. Next, using a field with a continuous slope, 
we describe our observations of chemotactic behavior in a broader regime 
than originally proposed. We explain this apparent
discrepancy from the Oosawa condition by the renormalization
of bare timescales
through the bacterial motion within the slope. Using these renormalized
timescales, we reconfirm the relevance of the Oosawa condition.

\section{Model}

We first introduced an internal state of a
cell that 
responds and adapts to the external concentration of the signal
chemical, and controls
the tumbling frequency. This internal variable is denoted by $u$, which might be, for
example, the concentration of some protein in the cell that responds to
the external chemical and controls the tumbling frequency.
This internal chemical responds to the concentration of the 
attractive chemical component (termed $S$ here) in the field. 
As the cell moves and the concentration of the external
chemical 
increases, the concentration $u$ increases
and returns to the original value, a process known as
adaptation~\cite{Koshland}. The simplest way to
have such adaptation dynamics is by introducing another internal chemical,
whose concentration is given by $v$, so that the change of concentrations
is governed by
\begin{eqnarray}
\frac{du}{dt}=f(u,v;S) , \nonumber \\
\frac{dv}{dt}=g(u,v).
\end{eqnarray}
We assume here that the fixed point solution $u^*,v^*$ given by
$f(u^*,v^*;S)=0,\ g(u^*,v^*)=0 $ is stable. If $f$ increases with 
$S$ and $u^*$ is independent of $S$, the response to $S$ and adaptation are
satisfied because $u$ increases with $S$ first, and then returns to the original $u^*$.
Here, when the solution $g(u,v)=0$ involves $u$ but not $v$,
the latter constraint is satisfied. For example,
$g(u,v)=\beta u v -\gamma v$ satisfies the condition, where
$\beta, \gamma$ are positive constants.
In this paper, we report our study of such a case, with
\begin{eqnarray}
\frac{du}{dt}=f(u,v;S)=S-\beta u v -\alpha u , \nonumber \\
\frac{dv}{dt}=g(u,v)=\beta u v -\gamma v .
\end{eqnarray}
\noindent
corresponding to the reaction process $S \rightarrow u$, $u+v
\rightarrow 2v$,
and degradation of $u$ and $v$. Another simple example,
originally introduced by Othmer~\cite{Othmer} is given by
$g(u,v)=(S- v)/\mu $, which gives the linear dynamics
\begin{eqnarray}
\frac{du}{dt}=\frac{S- (u + v)}{\eta } , \nonumber \\
\frac{dv}{dt}=\frac{S- v}{\mu }.
\end{eqnarray}
\noindent
with $\eta$ and $\mu$ as positive constants.  
We also briefly discuss the result of this model.  In both models, after
$S$ increases, $u$ first increases but later returns to the
$S$-independent fixed-point concentration given by 
$u^*=\gamma/\beta$ in the model eq.(2). 

We set the parameter values so that the
fixed-point solution $u^*,v^*$ is stable. In the model eq.(2), this
condition is given by $ \alpha \gamma < \beta S $.
In this case, following the increase (decrease) of
$S$, $u$ first increases (decreases) from $u^*$ and then returns 
to the original value exponentially with time and shows
a peak in time. The sensing time $\tau_s$ is estimated as the time to reach
this peak and is given by $ \sim 1/ 2S $. In contrast, the
adaptation time $\tau_a$ is estimated by the relaxation time towards the
fixed point, given by the eigenvalue of the linearized equation
of (2) around the fixed point, and is $ \sim 1/ \gamma $.
In this way, the internal timescales are represented by
the reaction parameters.

In general, the tumbling frequency is controlled by 
the internal state $u$, which we assume
is given by a continuous function of the concentration $u$.
The tumbling occurs randomly in experiments, whereas the rate changes in response to
the external signal $S$.  Although the speed could also change slightly
in bacteria, for simplicity we have changed only the rate of tumbling.
We assume that the cell moves with a constant speed, until it
changes direction, whose
probability (i.e., the rate of
tumbling) $1/\tau(u)$ is given by a function of $u$.
For $u=u^*$, we set the rate as $1/\tau^*$ and
assume that $\tau(u)$ is an increasing function for $u$.
As an example, we take the form
\begin{math}
\frac{1}{\tau(u)}=\frac{1.5-\tanh(\lambda (u-u_0))}{\tau^*}
\end{math} and choose the parameters so that
$\tau(u)$ approaches $2\tau^*$ for $u>u^*$, and 
$\tau^*/2$ for $u<u^*$, while $\lambda$ is set
sufficiently large to respond a change in $u$ by eq.(2).
This specific choice is not essential and 
the results we discuss are valid provided that the rate ($1/\tau(u)$) approaches a value 
sufficiently smaller than $1/\tau^*$ for a large $u$, and
sufficiently larger than $1/\tau^*$ for small $u$.
This change in the tumbling frequency is
consistent with experimental data~\cite{Berg1}.

Note that the normal tumbling rate $1/\tau^*$ is 
independent of the intracellular dynamics, so both are
independent of $\tau_s$ and $\tau_a$. This allows us to examine the
validity of the Oosawa condition.
Hereafter, we take the elements satisfying eq.(2) and insert them 
into a one-dimensional space, where the external concentration 
is given as $S(x)$, to examine
whether the elements assemble in a region with larger $S(x)$.

\section{Results on the Condition for Chemotaxis}
\subsection{Case 1: Step change in the chemical field}

\begin{figure}
\begin{center}
\scalebox{0.45}{\includegraphics{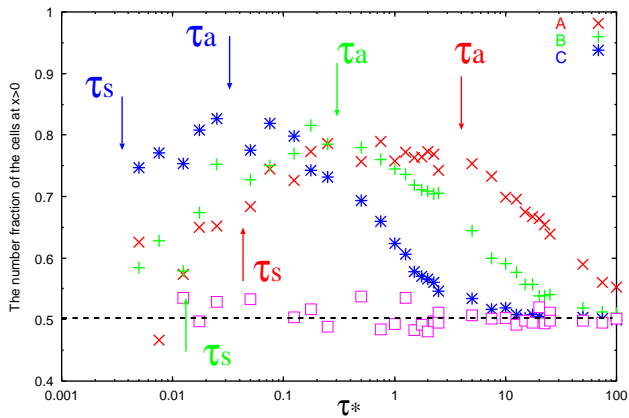}}
\caption{The number fraction of the cells 
located at $x \geq 0$ (i.e., in the $S$-rich field) are
plotted as a function of $\tau^*$.
The fraction is obtained from the temporal average 
for the timespans of 50,000 ($1 \leq \tau^*$), 200,000 ($0.1 \leq \tau^* < 1$),
and 300,000 ($\tau^* < 0.1$),
after the steady-state distribution is reached over 100 cells.
See Table~\ref{timeorder} for the parameters of the model eq.(2) 
with A, B, and C; the data without internal dynamics are also plotted 
$(\square)$ for reference.
    \label{step}}
\end{center}
\end{figure}

As a first example, we consider the case with a step-function external
field (i.e., $S(x)=S^+$ for $ x \geq 0$ and
$S(x)=S^-$ for $ x < 0$) and examine whether the cells assemble in the region
$x \geq 0$.
The number fraction of cells at $x \geq 0$, numerically obtained
for the steady state is plotted
in Fig.~\ref{step} by changing the parameter $\tau^*$.  In the simulation,
all cells were initially located at $x<0$.
The figure shows three examples of
different internal timescales $\tau_s$ and $\tau_a$,
and the data without
internal dynamics are plotted for reference.
Table.~\ref{timeorder} shows the time scales($\tau_s$ , $\tau_a$ ,
$\tilde{\tau_s}$(to be defined later)) for the three cases we used in the plot.

As shown, the fraction of cells for $x \geq 0$ peaks
as $\tau^*$ changes.  For all cases, the
most effective tumbling time $\tau_{peak}$ 
that gives a peak for the fraction  
lies at $\tau_s$
\mbox{\hspace{0.3em}\raisebox{0.4ex}{$<$}\hspace{-0.75em}\raisebox{-.7ex}{$\sim$}\hspace{0.3em}}
$\tau_{peak}$
\mbox{\hspace{0.3em}\raisebox{0.4ex}{$<$}\hspace{-0.75em}\raisebox{-.7ex}{$\sim$}\hspace{0.3em}}
$\tau_a$. All simulations for other parameters show that 
the elements assemble in the region $x \geq 0$, that is,
the chemotaxis works well for $\tau_s$ 
\mbox{\hspace{0.3em}\raisebox{0.4ex}{$<$}\hspace{-0.75em}\raisebox{-.7ex}{$\sim$}\hspace{0.3em}}
$\tau^*$  \mbox{\hspace{0.3em}\raisebox{0.4ex}{$<$}\hspace{-0.75em}\raisebox{-.7ex}{$\sim$}\hspace{0.3em}}
$\tau_a$.  These
results  confirm the Oosawa condition.
In contrast, as $\tau^*$ increases for $\tau* >\tau_a$ or
decreased for $\tau^* <\tau_s$, the fraction approaches $0.5$, implying
that chemotaxis is not possible.

\begin{table}[h]
\begin{ruledtabular}
\caption{The parameters and the timescales.  \label{timeorder}}
\begin{tabular}{|l||c|c||c|c|c|}
\multicolumn{1}{|c||}{Case} & $\alpha$ & $\gamma (=\beta/2)$ & $\tau_s$ & $\tau_a$ & $\tilde{\tau_s}$ \\
\hline \hline
 A &  $0.35$  &  $ 0.25$   &  $ 0.035$   &  $ 3$     &  $10 \sim 30$    \\ \hline
 B &  $3.5$   &  $ 2.5$    &  $ 0.015$   &  $ 0.3$   &  $1 \sim 4$      \\ \hline
 C &  $35.0$  &  $ 25.0$   &  $ 0.0035$  &  $ 0.03$  &  $0.2 \sim 0.5$  \\
\end{tabular}
\end{ruledtabular}
\end{table}

\subsection{Case 2: Chemical gradient with a constant slope}

Next, we consider the case with an external chemical concentration 
at a constant gradient (i.e.,
$S(x)=S_0+sx$).  Again, by initially
positioning  all cells in the region with small $x$ with low attractant
concentration, we can quantify whether cells move toward the larger $x$.
In this case, the cells with internal adaptive dynamics 
move toward the larger $x$ and stay there, regardless of their 
tumbling time $\tau^*$.

\begin{figure}
\begin{center}
\scalebox{0.45}{\includegraphics{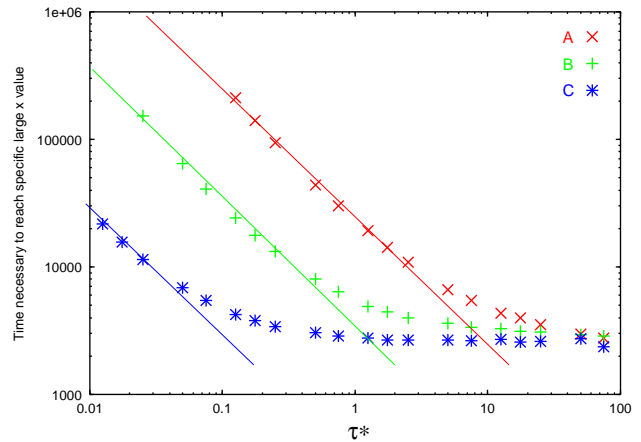}}
\caption{The time $T$ to reach a sufficiently large $x$ value, 
plotted as a function of $\tau^*$. 
The parameter values of intracellular dynamics are shown in Table~\ref{timeorder}.
The time $T$ was estimated when the center of 100 cells reached $x>1000$, 
after positioning 100 cells initially at $x<10$.
   \label{time}}
\end{center}
\end{figure}

The speed for climbing up the slope depends on $\tau^*$ and
the relation between $\tau^*$ and the internal timescales $\tau_s$ and
$\tau_a$.
To check the speed, we examined the time $T$ necessary to reach a specific
large $x$ value, as shown in Fig.~\ref{time}.
As $\tau^*$ becomes smaller, the time $T$ increases with $1/\tau^*$ for
$\tau^*< \tau^*_c$ with some critical value $\tau^*_c$ that depends on
$\tau_s$ and $\tau_a$.
In other words, efficient adaptive motion requires a threshold value
$\tau^*> \tau^*_c$. The next step was to examine whether
this range of $\tau^*$ satisfies the Oosawa condition given by
$\tau_s$ and $\tau_a$.

We noted that even if $\tau^* \gg \tau_a$, the cells can move to
a larger $x$ efficiently,
whereas the critical $\tau^*_c$ needed to increase $T \sim 1/\tau^*$ is
much larger than $\tau_s$.
This seemed to violate the original form of the Oosawa condition ( $\tau_s$ 
\mbox{\hspace{0.3em}\raisebox{0.4ex}{$<$}\hspace{-0.75em}\raisebox{-.7ex}{$\sim$}\hspace{0.3em}}
$\tau^*$ 
\mbox{\hspace{0.3em}\raisebox{0.4ex}{$<$}\hspace{-0.75em}\raisebox{-.7ex}{$\sim$}\hspace{0.3em}}
$\tau_a$ ).

\begin{figure}
\begin{center}
\scalebox{0.3}{\includegraphics{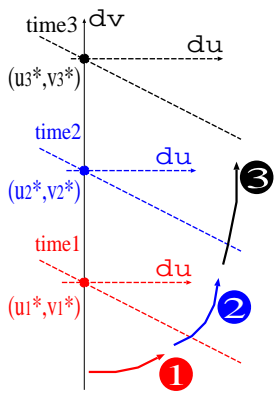}}
\caption{The orbit of $(u,v)$ as $S$ changes continuously. 
The large points are the equilibrium points $(u^*,v^*)$ for 
given $S$ values, whereas the dashed lines are stable manifolds at each time.
The curved lines with the arrow show the locus of $(u,v)$.
Before $(u,v)$ reaches the equilibrium point $(u_1^*,v_1^*)$ for $S$ value 
of the time 1, the equilibrium point shifts to $(u_2^*,v_2^*)$.
Accordingly, $(u,v)$ hardly reaches an equilibrium point.
    \label{orbit}}
\end{center}
\end{figure}

We note that the actual response and adaptation
times for cells moving in the
described concentration field are modified from those estimated from the
original
intracellular dynamics.
As the cell continuously moves and senses the change of the external concentration,
these timescales `renormalize'.
To examine this effect, we studied the intracellular dynamics more
closely, because
the values of $\tau_s$ and $\tau_a$ may depend
on the dynamics of the internal system $(u,v)$ and $S$ is changing continuously with time
according to the cell's motion.
In our model, as $S(x)$ continued to change, the equilibrium point of $(u,v)$  also 
changed, which invalidated the baseline $u^*,v^*$ to give $\tau_s$ and $\tau_a$ obtained by
the linear approximation method (Fig.~\ref{orbit}). 

Instead, by tracking the time series of $u$ and measuring the
time for the peak and relaxation time to $u^*$,
we estimated the renormalized values $\tilde{\tau_s}$ and
$\tilde{\tau_a}$. First, as a cell moves to
a region with a higher signal concentration $S$, the
relaxation to the original $u^*$ value hardly occurs because, as 
$S$ increases, the increase in $u$ occurs before relaxing to $u^*$.
Hence, the renormalized $\tilde{\tau_a}$ is infinite or at least $\gg \tau_a$.
Accordingly, one part of the Oosawa condition $\tau^*$ 
\mbox{\hspace{0.3em}\raisebox{0.4ex}{$<$}\hspace{-0.75em}\raisebox{-.7ex}{$\sim$}\hspace{0.3em}}
$\tau_a$ is always satisfied.

\begin{figure}
\begin{center}
\scalebox{0.45}{\includegraphics{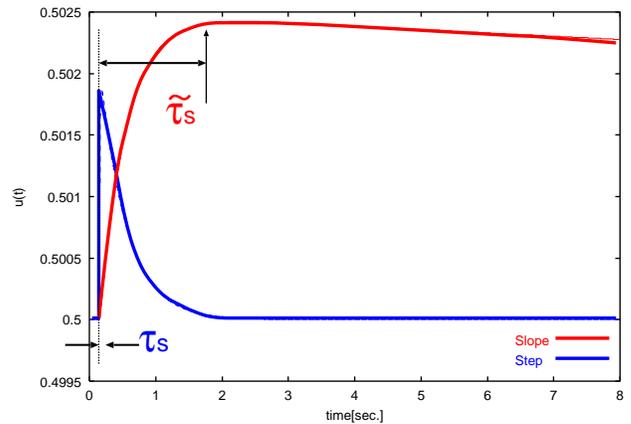}}
\caption{ The time series of $u$ for the step chemical field $S(x)$ 
(blue) and the gradient with a constant slope (red).
The series uses parameter values for the case B in Table~\ref{timeorder}, 
where the fixed-point value $u^*$ is 0.5.
The response to the change of $S$ at $t=0.15$ is plotted.
     \label{dynamics}}
\end{center}
\end{figure}

The sensing time also changes from $\tau_s$ when cells
are placed in a continuously changing field. 
Although $u$ does not relax to the original $u^*$, 
it increases with $S$ and decreases slowly, as
shown in Fig.~\ref{dynamics}.  
The renormalized sensing time $\tilde{\tau_s}$ can be estimated by the time
needed to reach the maximal value.
The renormalized $\tilde{\tau_s}$ increases from $\tau_s$ 
as cells move before $u$ reaches the original peak for a given concentration of $S$
because the fixed point of $(u^*,v^*)$ continuously changes, 
as depicted in Fig.~\ref{orbit}.

Now, $\tilde{\tau_s}$, the threshold
beyond which the (renormalized) Oosawa condition $\tilde{\tau_s}$ 
\mbox{\hspace{0.3em}\raisebox{0.4ex}{$<$}\hspace{-0.75em}\raisebox{-.7ex}{$\sim$}\hspace{0.3em}}
$\tau^*$ is satisfied, is expected to give a criterion for chemotaxis.
We compared $\tilde{\tau_s}$ with $\tau^*_c$ , the critical $\tau^*$ value, 
to show the increase of $T \sim 1/\tau^*$ (see Fig.~\ref{time}) and found that
the two values agree with each other.
We note that
$\tau^*$ 
\mbox{\hspace{0.3em}\raisebox{0.4ex}{$>$}\hspace{-0.75em}\raisebox{-.7ex}{$\sim$}\hspace{0.3em}}
$\tau^*_c (\sim \tilde{\tau_s})$ is the only condition for chemotaxis,
because $T$ maintains a constant value over a wide range for $\tau^*$
\mbox{\hspace{0.3em}\raisebox{0.4ex}{$>$}\hspace{-0.75em}\raisebox{-.7ex}{$\sim$}\hspace{0.3em}}
$\tau^*_c$. Thus, the (renormalized) Oosawa condition
$\tilde{\tau_s}$ 
\mbox{\hspace{0.3em}\raisebox{0.4ex}{$<$}\hspace{-0.75em}\raisebox{-.7ex}{$\sim$}\hspace{0.3em}}
$\tau^*$  
\mbox{\hspace{0.3em}\raisebox{0.4ex}{$<$}\hspace{-0.75em}\raisebox{-.7ex}{$\sim$}\hspace{0.3em}}
$\tilde{\tau_a}$ correctly estimates the condition for chemotaxis.
Finally, we note that the renormalized values
$\tilde{\tau_s}$ and $\tilde{\tau_a}$ are
independent of the slope of the concentration of $S$, and that this condition
gives a criterion for chemotaxis for any gradient in the concentration.

\begin{figure}
\begin{center}
\scalebox{0.45}{\includegraphics{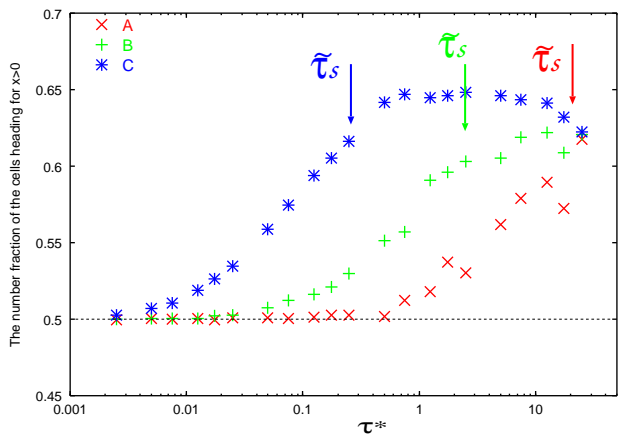}}
\caption{ The number fraction of cells moving toward the larger $x$, 
(i.e., for large $S(x)$) at $T=100$.
The parameters for intracellular dynamics are shown in Table~\ref{timeorder}, in which
all conditions are the same as those in Fig~\ref{time}.
The fraction 0.5 indicates the absence of chemotaxis in short-term behavior.
   \label{Slop}}
\end{center}
\end{figure}

The time $T$, estimated above as the value necessary to reach 
the region with higher attractant concentration,
characterizes the ability of chemotaxis over the long term, as mentioned by
Clark and Grant~\cite{Clark}. In contrast, the condition for
efficient chemotaxis in a shorter-term is different in general.
In our case, however, the chemotactic behaviors in the short and long term are not independent,
but are related through intracellular dynamics.
We also examined the short-term chemotactic behavior
by measuring the number fraction of cells
that moved toward the higher attractant concentration at $T=100$,
starting from a random distribution of cells.
The fraction is plotted as a function of $\tau^*$ in Fig.~\ref{Slop},
for three examples with different internal timescales.
Here again, if $\tau^*$  
\mbox{\hspace{0.3em}\raisebox{0.4ex}{$>$}\hspace{-0.75em}\raisebox{-.7ex}{$\sim$}\hspace{0.3em}}
$\tilde{\tau_s}$, the fraction is much larger
than 0.5, whereas
for $\tau^* \ll \tilde{\tau_s}$, the fraction is about $0.5$, indicating that the
cells move
randomly. Hence, the Oosawa condition for chemotaxis is also valid
for short-term behavior. Because the internal adaptive dynamics satisfy
the renormalized Oosawa condition, we conclude that
chemotaxis works well over both the short and long term.

\section{Discussion}

We have presented a model of chemotaxis that includes 
internal adaptive dynamics and have shown that the chemotactic behavior appears  
when the Oosawa condition is
satisfied, that is, when the timescale of tumbling is
greater than the signal-response time and smaller than the time for the adaptation.

Our conclusion about the condition for chemotaxis
applies generally for a system with intracellular adaptive dynamics.
For example, we have also studied the
linear model given by eq.(3)~\cite{Othmer}. By changing the
internal timescales $\tau_s$ and $\tau_a$, we have examined the validity
of the Oosawa condition for chemotaxis. 
Whether chemotaxis works efficiently is determined by the Oosawa condition for both
the chemical concentration field of a step-function and a constant slope,
and both for short- and long-term behavior.
 
To perform chemotaxis, the cell's internal dynamics have to sense and respond to changes in the
the external chemical concentration. The timescale of sensing should be
smaller than that of the tumbling frequency to induce an effective
response.  Otherwise, random walks are repeated before the response occurs.
Thus, the condition $\tau_s$ 
\mbox{\hspace{0.3em}\raisebox{0.4ex}{$<$}\hspace{-0.75em}\raisebox{-.7ex}{$\sim$}\hspace{0.3em}}
$\tau^*$ is essential
to the short-term response.  In contrast, response of the tumbling frequency without adaptation 
causes the long-term behavior of cells to be represented by a
random walk, so that chemotaxis is not possible.  
This implies the need for intracellular adaptive dynamics (i.e., relaxation to the
original value).  However, to induce an effective
response to external changes, the tumbling should occur before the relaxation
is completed.  Hence, the condition $\tau^*$  
\mbox{\hspace{0.3em}\raisebox{0.4ex}{$<$}\hspace{-0.75em}\raisebox{-.7ex}{$\sim$}\hspace{0.3em}}
$\tau_a$ is required.  
Our numerical results suggest that the Oosawa condition $\tau_s$ 
\mbox{\hspace{0.3em}\raisebox{0.4ex}{$<$}\hspace{-0.75em}\raisebox{-.7ex}{$\sim$}\hspace{0.3em}}
$\tau^*$ 
\mbox{\hspace{0.3em}\raisebox{0.4ex}{$<$}\hspace{-0.75em}\raisebox{-.7ex}{$\sim$}\hspace{0.3em}}
$\tau_a$
is both necessary and sufficient to explain the chemotactic behavior.

Clark and Grant~\cite{Clark} recently reported the conditions for
the internal response function $R(t)$ to show chemotaxis.
In general, the conditions for
short- and long-term chemotaxis differ, but by
imposing some type of optimization to balance the two behaviors, they were able to
obtain a certain condition for the response function, which implies the existence of
a form of memory in the autocorrelation function.  Our work answers the
question about why chemotaxis requires dynamics.
We can estimate the response function $R(t)$ from our model,
which shows that the proper response function in their sense is
obtained when the Oosawa condition is satisfied.
Similarly, we can define the Oosawa condition 
in the response function $R(t)$ of Clark and Grant.
The sensing time $\tau_s$ is estimated from the response function 
as $R(\tau_s)=0$, and their solution satisfies $\tau_s \sim \tau^*$, 
whereas $\tau_a$ is much larger than $\tau^*$.

Although we recognize the importance of their paper~\cite{Clark}, 
we believe some controversy remains about the 
optimization condition.
We have assumed the existence of intracellular dynamics and
that these apply to adaptation after considering the natural
properties of intracellular dynamics~\cite{Koshland}.
By assuming a class of intracellular dynamics, we have shown that 
chemotaxis occurs in both short- and long-term behavior 
without requiring an optimization condition, provided that the Oosawa condition
is satisfied.
Because short- and long-term chemotactic behaviors are
satisfied simultaneously, one expects that cells can choose such internal
adaptation dynamics
that enable efficient chemotaxis for a variety of external conditions.
Here we have demonstrated chemotaxis in the two cases: with a constant slope and with a stepwise change.
Because most environmental conditions can be represented by the combination of
these two cases, the Oosawa condition
gives a criterion for chemotaxis in general. Furthermore,
we have confirmed that chemotaxis works when
spatial--temporal noise is added in the 
external environment.

Because the optimal condition in the
internal adaptive dynamics depends on the external motility (tumbling time),
the validity of the Oosawa condition can be checked experimentally.
The table~\ref{internal} shows the timescales estimated from the data obtained for the wild-type and
mutant cells of \textit{ E. coli} shown in \cite{Berg1,Berg2} (see Fig.3 and 7 \cite{Berg1}),
which supports the validity of the condition.
In addition, the data in \cite{Alon} suggest that both $\tau^*$ and $\tau_a$
change in mutants whereas the relationship is maintained $\tau^*$ 
\mbox{\hspace{0.3em}\raisebox{0.4ex}{$<$}\hspace{-0.75em}\raisebox{-.7ex}{$\sim$}\hspace{0.3em}}
$\tau_a$ if the mutant does not lose its chemotactic ability.
In general, different organisms have different timescales for tumbling and
internal timescales($\tau_s$ and $\tau_a$).
It will be important to check the validity of the condition in different organisms.

\begin{table}[bp]
\begin{ruledtabular}
\caption{The internal time scales of \textit{E. coli.} (seconds)\label{internal}}
\begin{tabular}{|l||c|c|c|}
\multicolumn{1}{|c||}{Cell} & $\tau^*$ & $\tau_s$ & $\tau_a$  \\
\hline \hline
 Wild-type            &  $1.5$   &  $ 0.4$  &  $ 4$ \\ \hline
 $che RB^{\Delta} $   &  $2.3$   &  $ 0.6$  &  $ 3$ \\ \hline
 $che Z $             &  $12$    &  $ 2.6$  &  (larger than $ 15$)  \\
\end{tabular}
\end{ruledtabular}
\end{table}

Living organisms have many reactions with various timescales, which, when combined, 
cause adaptive functions to emerge. The proper use of
different timescales is important when generating adaptive output behavior
in response to changes in external conditions. The process of chemotaxis that we have discussed is one of the
simplest mechanisms that takes advantage of the timescale differences,
whereas the condition for different timescales
can be generalized in a more complex reaction-network system, 
and in intercellular interactions\cite{inoue-future}.

The authors wish to thank
Koichi Fujimoto, Shuji Ishihara, Katsuhiko Sato, Masashi Tachikawa, 
Satoshi Sawai, Naoto Kataoka, and
Akihiko Nakajima for useful discussion and insightful comments.

\end{document}